**Magnetic properties of the Heavy Fermion Antiferromagnets YbNiAl and YbPtAl**

J. Diehl[*], H. Davideit[*], S. Klimm[*], U. Tegel[+], C.Geibel[+], F.Steglich[+] and S. Horn[*]

[*]Lehrstuhl für Experimentalphysik II, Uni Augsburg, 86135 Augsburg, Germany.

[+]Institut für Festkörperphysik, TH Darmstadt, 64289 Darmstadt, Germany.

**abstract**

Measurements of electrical resistivity and Hall effect as a function of magnetic field on the Heavy Fermion Systems YbNiAl and YbPtAl are presented. Both compounds order antiferromagnetically and show metamagnetic behavior in a magnetic field. Scaling behavior of the magnetoresistance above $T_N$ suggests that the paramagnetic regime for YbNiAl can be described in terms of a single ion Kondo effect.

**Keywords :** Yb compounds, heavy fermions, hall effect, magnetoresistance

**contact address :** Jürgen Diehl
Institut für Physik
Universität Augsburg
Memmingerstr. 6
86135 Augsburg
Germany
Tel: +49 821 / 5977-453
Fax +49 821 / 5977-411
e-mail : diehl@physik.uni-augsburg.de

There are several examples of metallic Yb-compounds which show the typical heavy fermion properties found also in Ce- and U-systems [1]. This is not surprising since the $Yb^{3+}$-ion can be considered to be the hole analogue of $Ce^{3+}$. Here we present results of electronic transport measurements on two new heavy fermion Yb-compounds; i. e. YbNiAl and YbPtAl which display a coefficient of the electronic specific heat of $\gamma$ = 350 mJ/mol$K^2$ and $\gamma$ = 200 mJ/mol$K^2$, respectively [2]. YbNiAl crystallizes in the hexagonal ZrNiAl structure ($Fe_2P$-Type) while YbPtAl has the orthorhombic ε-TiNiSi structure. Polycrystalline material of these two compounds was prepared in closed molybdenum crucibles under Argon atmosphere and characterized by x-ray diffraction [2]. The electrical resistivity and the Hall effect were measured in magnetic fields up to 14.5 T, using either a standard four probe AC lock-in technique or the Van der Pauw method. Measurements were performed in the temperature range 0.3 K < T < 300 K. Magnetization measurements were carried out in a SQUID magnetometer in the temperature range 1.7 K < T < 400 K.

The magnetic susceptibility of YbNiAl follows a Curie-Weiss law between T = 400 K and T = 30 K, with an effective moment of 4.4 $\mu_B$ per Yb-atom, and a Curie-Weiss temperature $\Theta_{CW}$ = - 35 K. At $T_N$ = 2.8 K a cusp in $\chi(T)$ is observed which is shifted to lower temperature with increasing field, indicating antiferromagnetic order (fig. 1). The cusp is suppressed completely in the investigated temperature range by a magnetic field of 4T. For YbPtAl an effective moment of 4.5 $\mu_B$ per Yb-atom and $\Theta_{CW}$ = - 60 K is inferred from a Curie-Weiss law, which is valid between 400 K and 150 K. A cusp in $\chi(T)$ observed at $T_N$ = 5.8 K indicates antiferromagnetic order, but toward lower temperature $\chi(T)$ increases again (fig. 1). A magnetic field of B = 0.3 T induces a steep increase in $\chi(T)$ below 3 K and a field of 0.5 T is sufficient to completely suppress the cusp observed in small fields.

The electrical resistivity of both compounds is shown in fig. 2, together with the non magnetic reference compound LuNiAl, which is isostructural to YbNiAl. For YbNiAl an initial decrease of ρ(T) upon cooling from room temperature is followed by a negative temperature characteristic down to about 3 K, indicative of Kondo scattering. The antiferromagnetic order at $T_N$ = 2.8 K causes a rapid decrease of ρ(T) at that temperature. YbPtAl exhibits an almost constant ρ(T) between T = 300 K and T = 150 K. A decrease of ρ(T) below 150 K is indicative of the presence of crystal field effects. After going through a minimum at T = 35 K and a maximum at 10 K there is a steep decrease below $T_{N1}$ = 5.8 K marking the onset of antiferromagnetism. In the inset of fig. 2 a second sharp drop of ρ(T) is visible at $T_{N2}$ = 1.3K, presumably signalling a transition into a different magnetic state. The magnetization M(B) measured at T = 2 K, i. e. in the antiferromagnetic phases, shows an inflection point at B = 2 T for YbNiAl and at B = 0.5 T for YbPtAl. At the same fields ρ(B) shows a steep decrease as a function of field (fig. 3). Similar behavior has been observed, e. g. for TmSe [3], and has been interpreted as a transition from an antiferromagnet to an induced ferromagnet.

The resistivity ρ as a function of applied magnetic field (transverse magnetic resistance) is shown in fig. 4 for various temperatures below and above $T_N$ for YbNiAl. Below $T_N$ a sharp decrease in ρ(B) is observed after the initial increase which, as has been demonstrated in fig. 3, coincides with an inflection point in M(B). We interpret the sharp drop in ρ(B) below $T_N$ as a strong reduction of magnetic scattering on entering the polarized state. Above $T_N$, ρ(B) is always negative with its slope decreasing with increasing temperature (fig. 4). A point of inflection is observed up to temperatures of 5 K. The shape of these curves suggests an interpretation in terms of a single ion Kondo effect. In this case one would expect scaling behavior of the magnetoresistance if plotted vs B/(T + T*), were T* is a measure for the Kondo temperature. We find excellent scaling behavior for T* = 3K, which is shown in the inset of fig. 4, for $\rho$(B)-curves between 5 K and 20 K.

The isothermal Hall resistivity $\rho_{xy}$ of YbNiAl is shown in fig. 5. At T = 2.0 K the initial slope $d\rho_{xy}/dB$ reaches an maximum of $R_H = + 14.5 \cdot 10^{-9}$ $m^3/C$. For B > 2 T, $\rho_{xy}$ is strongly non linear. A maximum is followed by a dramatic drop in $\rho_{xy}$ for $T < T_N$. This behavior tracks that of the magnetoresistance and results presumably from a strong suppression of magnetic scattering at the transition from antiferromagnetism to induced ferromagnetism. For T = 0.4 K the slope of $\rho_{xy}$ at high fields (B > 5 T) takes a constant negative value which is equal to that of the nonmagnetic reference compound LuNiAl. This suggests that for T = 0.4K magnetic scattering is completely suppressed for B > 5 T. For T > 200K the Hall coefficient of YbNiAl and of nonmagnetic LuNiAl take the same value. Above $T_N$ a broad maximum is observed in $\rho_{xy}$, which shifts to higher fields with increasing temperature. A completely linear $\rho_{xy}$ is observed in the available field range for T > 50 K. For the temperature range, in which magnetization and magnetoresistance measurements exist, $\rho_{xy}$ can, at least qualitatively, be described by $\rho_{xy} \sim \rho_{xx} \cdot M$. Such a behavior is expected, if the Hall resistance is dominated by skew scattering [4].

The magnetoresistance of YbPtAl (not shown) shows a qualitatively similar behavior to that of YbNiAl, but no scaling is found in a plot $\rho_{xx}$ vs B/(T + T*) for temperatures above $T_N$ = 6 K. A possible explanation for the failure of scaling is the existence of crystal field effects, which are clearly visible in $\rho$(T) of YbPtAl, but are not found for YbNiAl. The Hall resistivity $\rho_{xy}$ of YbPtAl shows a saturation behavior at low temperatures and becomes very small for T < 2 K for the whole field range. This behavior is consistent with the conclusion drawn from magnetization and magnetoresistance measurements (fig. 3), that ferromagnetism is induced in this compound already for small magnetic fields. This would suppress skew scattering and $\rho_{xy}$ would correspond to the normal Hall effect of the nonmagnetic reference compound.

In summary, magnetotransport and magnetization measurements show that YbNiAl and YbPtAl show metamagnetic behavior in an applied magnetic field below the Neél temperature. For YbNiAl the magnetotransport measurements can be described by a single ion Kondo behavior with a characteristic temperature of $T^* = 3$ K.

**Acknowledgment**

We would like to thank G.R. Stewart for the use of his SQUID magnetometer.

[1] Z. Fisk and M. B. Maple, Journal of Alloys and Compounds **183** (1992) p.303

[2] C. Schank et.al., submitted to ICM 1994

[3] P. Haen et.al. in 'Valence Instabilities and Related Narrow-Band Phenomena', ed. R.D. Parks (1977), p.495

[4] A. Fert and P.M. Levy, Physical Review B **36** (1987) p.1907

Fig.1. DC suszeptibility of YbNiAl (left) and YbPtAl (right). Magnetic fields as indicated in the figure.

Fig.2. Resistivity of YbNiAl, YbPtAl and LuNiAl. The inset shows the low temperature range on an expanded temperature scale.

Fig.3. DC magnetization and resistivity as a function of magnetic field B for YbNiAl (left) and YbPtAl (right) at T=2.0K.

Fig.4. Isothermal resistivity curves as a function of B for YbNiAl and LuNiAl at temperatures as indicated. The inset shows the magnetoresistance plotted vs $B/(T+T^*)$ for temperatures in the range $5K < T < 20K$.

Fig.5. Isothermal Hall resistivity curves of YbNiAl for various temperatures. The Hall coefficient of YbNiAl and LuNiAl coincides for $T > 200K$.

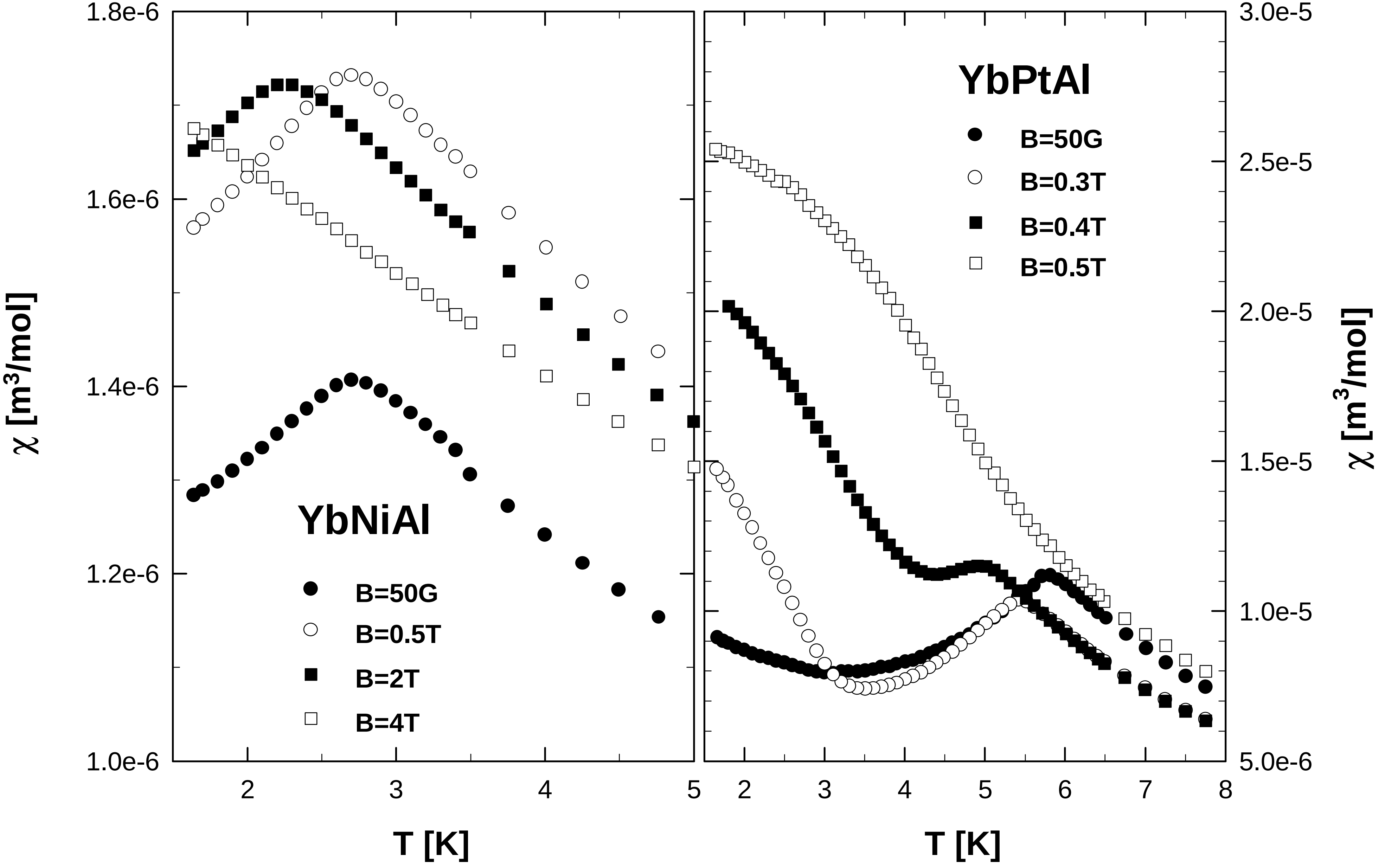

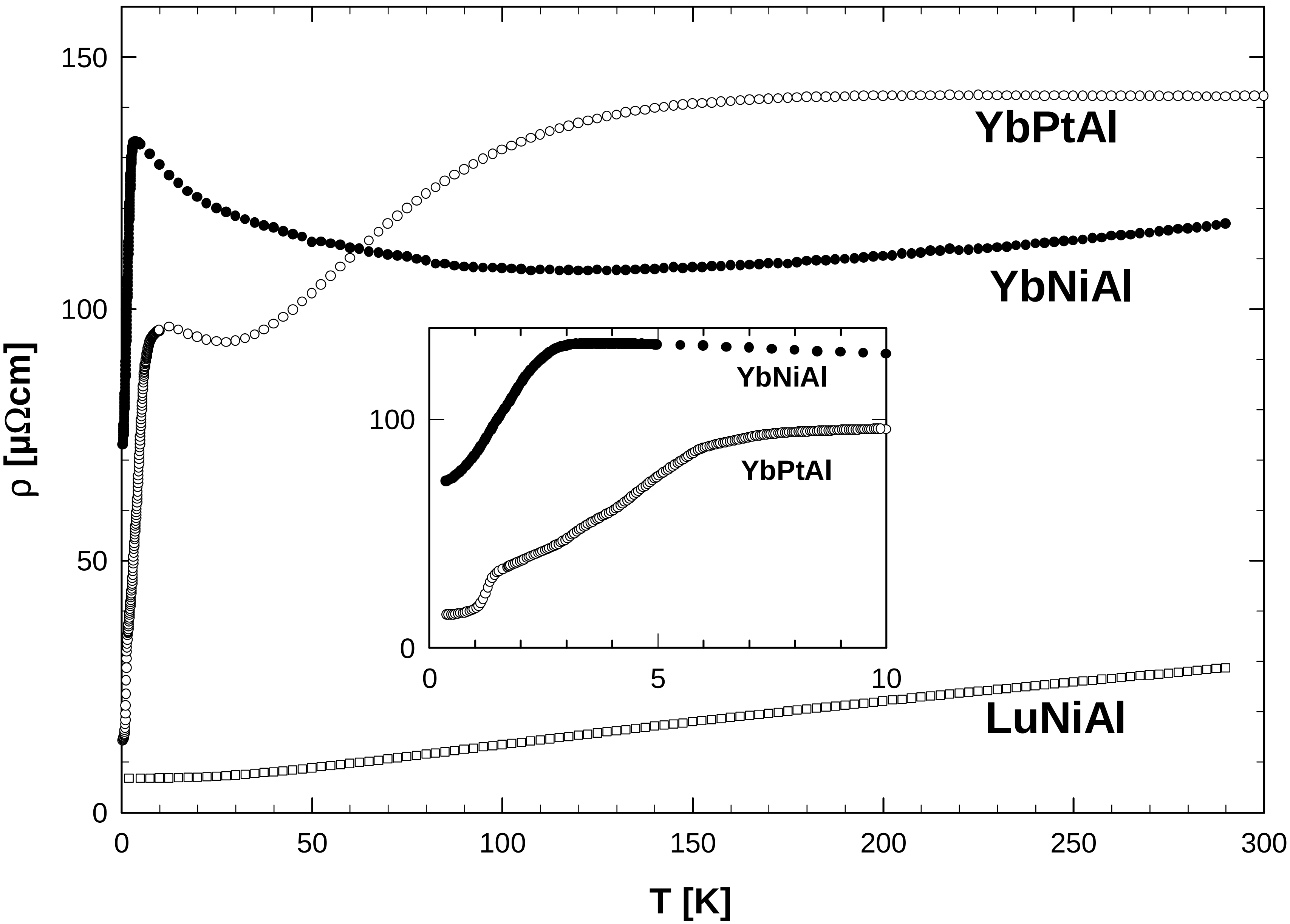

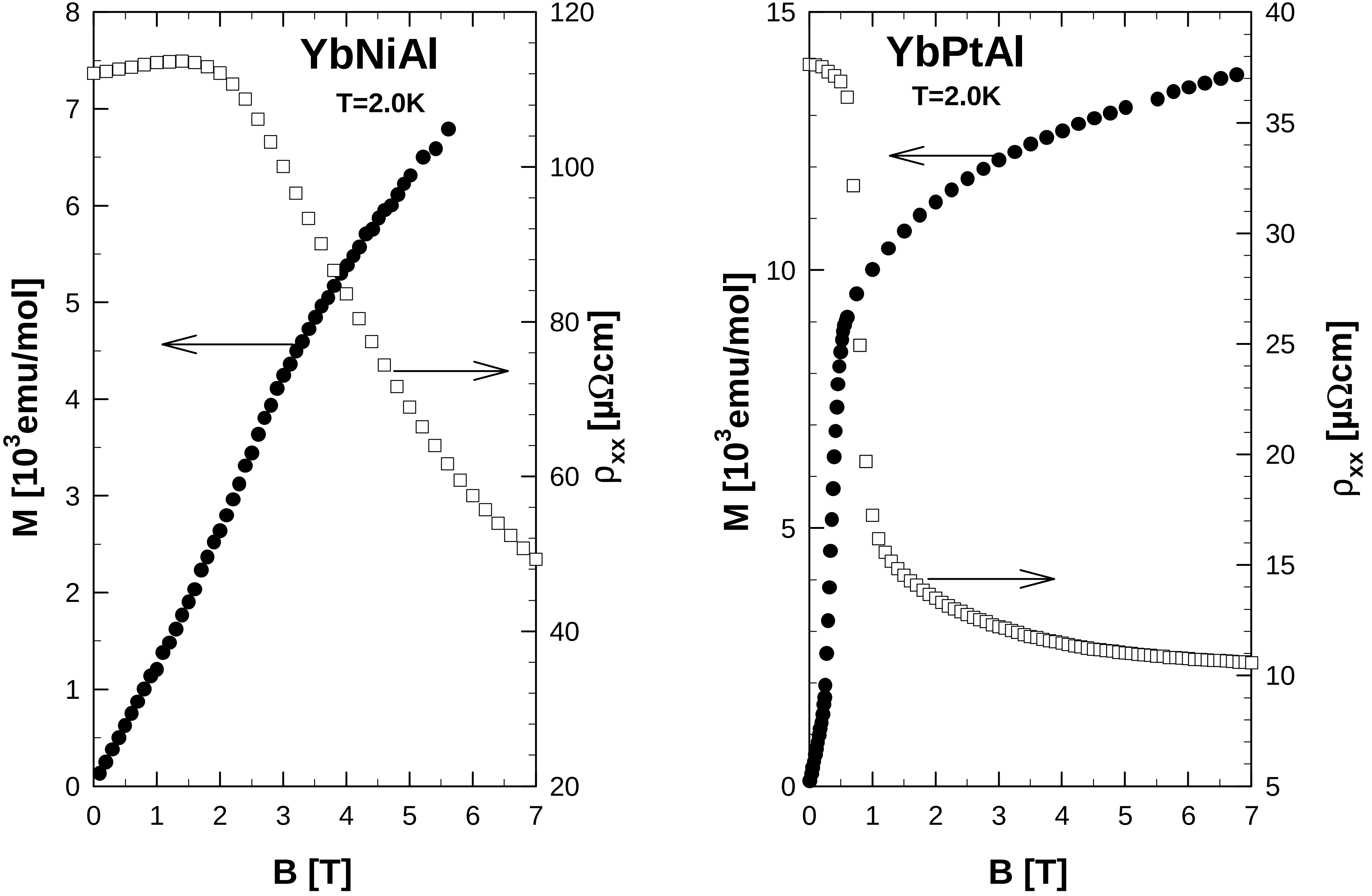

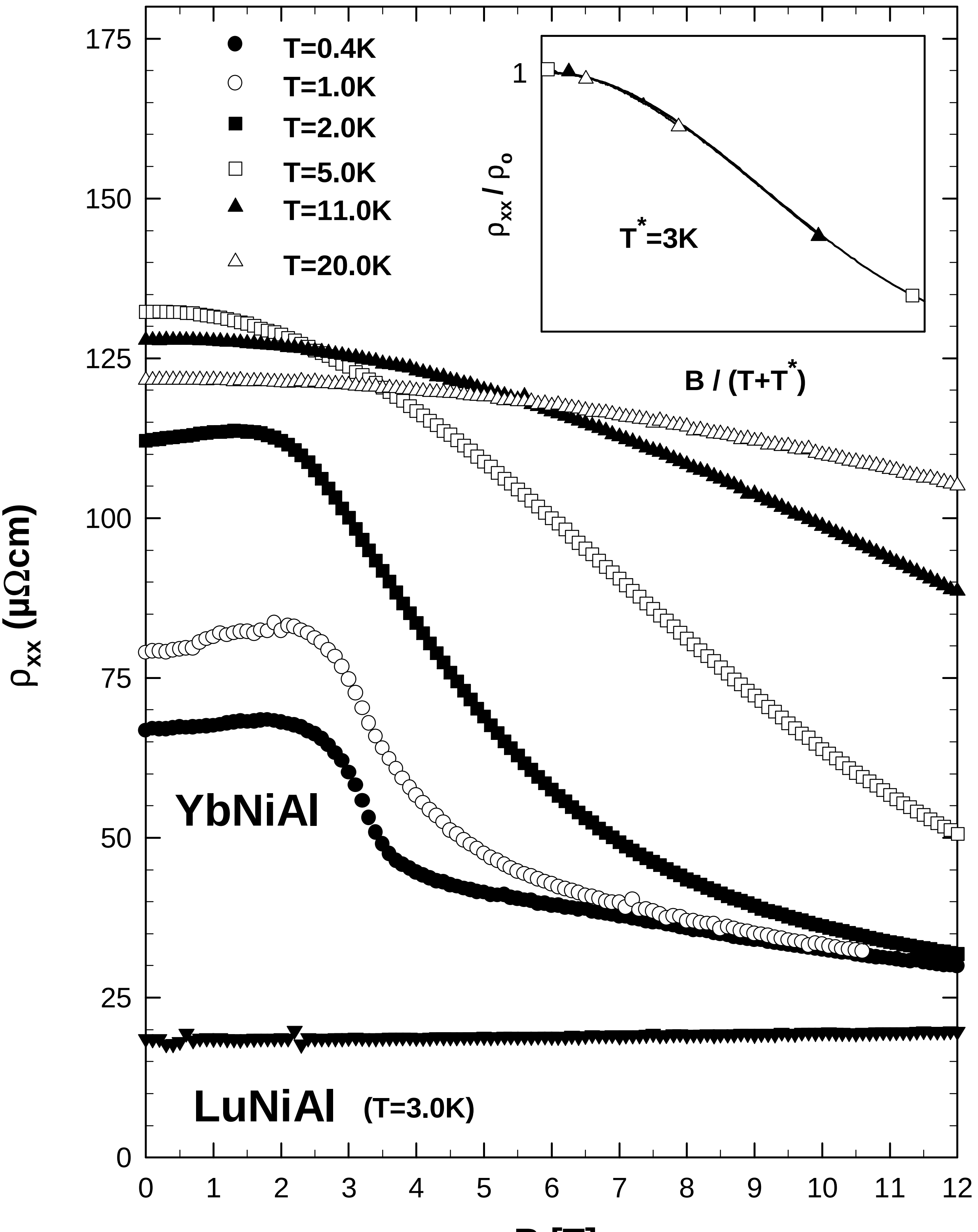

T=0.4K
T=1.0K
T=2.0K
T=5.0K
T=11.0K
T=20.0K
YbNiAl
LuNiAl (T=3.0K)
ρxx (μΩcm)
B [T]
ρxx / ρo
T*=3K
B / (T+T*)
1
0
25
50
75
100
125
150
175
0 1 2 3 4 5 6 7 8 9 10 11 12

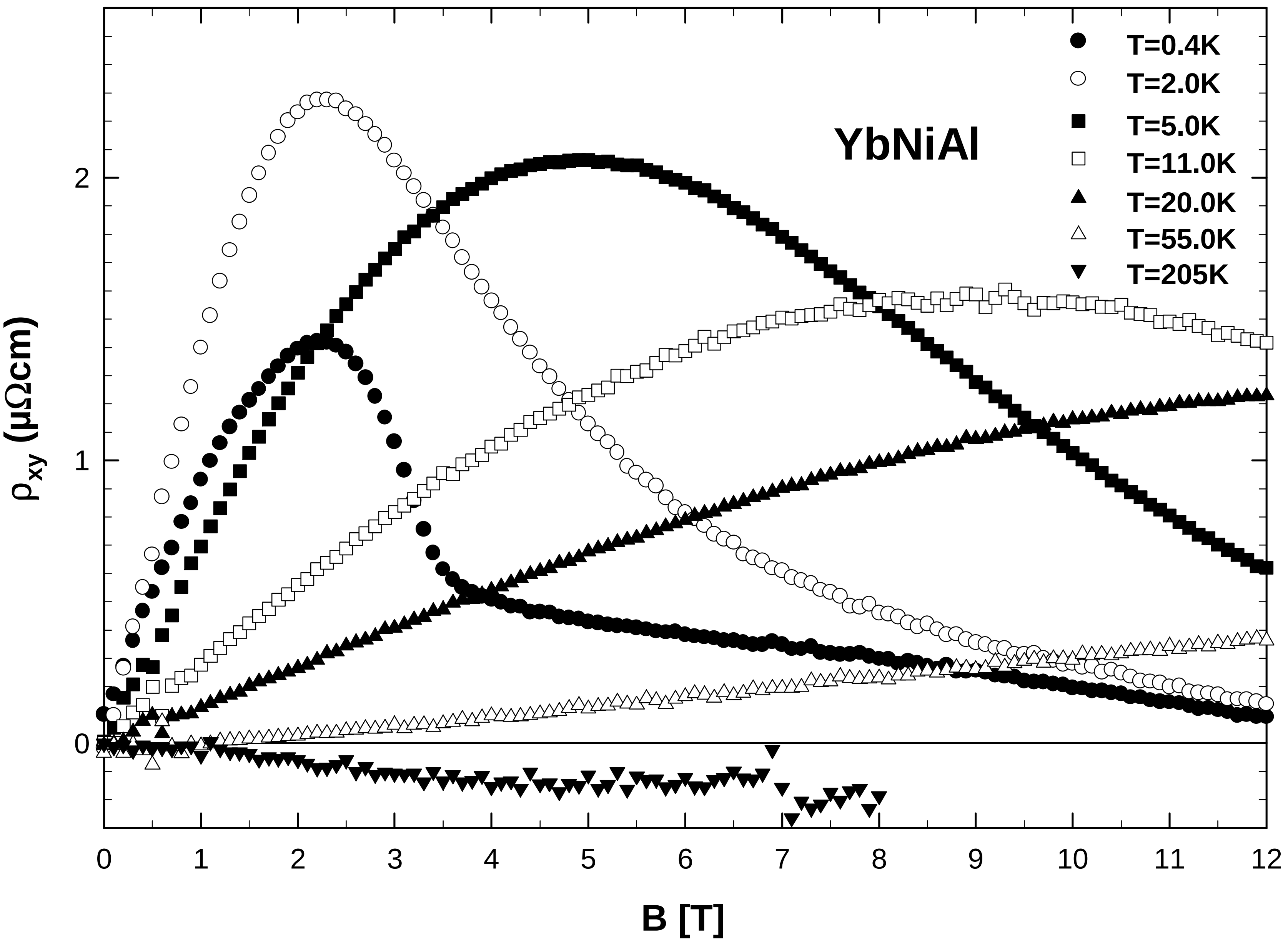